\documentclass{article}

\usepackage{arxiv}

\usepackage[utf8]{inputenc} 
\usepackage[T1]{fontenc}    
\usepackage{hyperref}       
\usepackage{url}            
\usepackage{booktabs}       
\usepackage{amsfonts}       
\usepackage{nicefrac}       
\usepackage{microtype}      
\usepackage{graphicx}
\graphicspath{{figures/}}
\usepackage{natbib}
\usepackage{doi}

\usepackage{siunitx}
\DeclareSIUnit\angstrom{\text {Å}}
\usepackage{tabularx}
\usepackage{authblk}
\usepackage{makecell}

\title{On the Robustness of Machine Learning Models in Predicting Thermodynamic Properties: a Case of Searching for New Quasicrystal Approximants}

\date{October, 2024}	

\author[1, 2]{Fedor S. Avilov}
\author[1]{Roman A. Eremin}
\author[1]{Semen A. Budennyy}
\author[1, *]{Innokentiy S. Humonen}

\affil[1]{AIRI, Moscow, Russia}
\affil[2]{National University of Science and Technology MISIS, Moscow}
\affil[*]{Corresponding author: Innokentiy S. Humonen, humonen@airi.net}

\renewcommand{\shorttitle}{\small{Robustness of ML Models in Predicting Thermodynamic Properties: a Sc-X case}}

\begin{document}

\maketitle
\small
\tableofcontents

\newpage

\begin{abstract}

Despite an artificial intelligence-assisted modeling of disordered crystals is a widely used and well-tried method of new materials design, the issues of its robustness, reliability, and stability are still not resolved and even not discussed enough. To highlight it, in this work we composed a series of nested intermetallic approximants of quasicrystals datasets and trained various machine learning models on them correspondingly. Our qualitative and, what is more important, quantitative assessment of the difference in the predictions clearly shows that different reasonable changes in the training sample can lead to the completely different set of the predicted potentially new materials. We also showed the advantage of pre-training and proposed a simple yet effective trick of sequential training to increase stability.
    
\end{abstract}


\section{Introduction}
Over the last decade, artificial intelligence (AI) has become a powerful tool for various tasks in numerous fields, such as natural language processing (NLP) \cite{owid-brief-history-of-ai, zhou2023comprehensive}, computer vision (CV) \cite{owid-brief-history-of-ai, haenlein2019brief}, natural sciences \cite{han2023revolutionizing, ivanenkov2023hitchhiker}, \textit{etc}. 
With the advent of neural networks, which are, unfortunately, uninterpretable black boxes \cite{alwosheel2021did}, the scientific and industrial communities began to pay attention to the issues of stability and robustness \cite{hamon2020robustness, kejriwal2024challengersjoshoconnorzendaya}.
First, let us define what we call 'robustness' hereafter.
It is known, that negligible changes in the input data (both for training model or inference model) can provoke uncontrolled changes in the output of AI.
Since the AI-assisted solutions affect real world decisions, this output, in turn, can lead to erroneous, costly, unethical, or even dangerous results.
Thus, the robustness of machine learning (ML) algorithms characterize absence of the aforementioned situations and seems to be one of the key issues in the design of AI-assisted systems.
The similar problems were discussed in detail by researchers in the NLP \cite{10.1145/3593042, wang2021measure} and CV \cite{LIU2023175, li2024survey} domains, but it can certainly arise wherever neural networks are used, including materials science. 

Indeed, AI fever has not spared materials science either: ML has become a solid part of the pipeline for searching for new materials by high-throughput thermodynamic stability assessments within the vast chemical search spaces. 
The most recent approaches in the field are resting upon testing chemical modifications of existing materials through their compositional/chemical modifications \cite{merchant2023scaling,Chen2024AcceleratingCM} instead of generation of crystal structures from scratch. 
Despite the fact that many modern computational datasets, such as that of \textit{Materials Project} \cite{jain2013commentary} and \textit{AFLOWLib} \cite{CURTAROLO2012227}, can be used to pre-train AI models for their subsequent applications, fine-tuning stage or even more sophisticated routines (e.g., active learning \cite{merchant2023scaling, yuan2023active}) are generally required for disordered materials. 
In such approaches, the more complex crystal structure, the more complex the target search space. 
For defects in 2D materials modelled within the supercell approach \cite{huang2023unveiling}, the authors provided combinations of structured composition/configuration spaces at low defect contents and random structures subsamples at higher ones. 
Nevertheless, it was recently demonstrated \cite{eremin2024perovskites} that physicochemical intuition can provide an acceptable quality of model predictions at low computational costs. 
Thus, training data sampling not only can play a major role in most effectively combining exhaustive and high-precision (such as density functional theory) calculations and fast screening data-driven approaches, but is also a point for potential instability: small errors and inaccuracies in the composition of the training dataset can lead to unnecessary and unsuccessful resource-intensive
ab-initio computational and expensive real-life experiments.

In our recent research \cite{eremin2022hybrid}, we addressed thermodynamic stability assessments for the disordered Sc-rich complex (\textit{ca}. 140 atoms/cell) intermetallics by combining density functional theory calculations and data-driven approaches.
During the data collection, there were obtained considerable differences in thermodynamic favorability of defects of certain types.
Despite this fact, all the data collected was used for model training.
This allowed graph neural networks (GNN) and descriptor-based ML models to gain sufficient generalization ability. 
In turn, at the inference stage, the complete composition/configuration space was built only for the thermodynamically favored defects resulting in an appropriate complexity of target search space. 
Thus, the peculiarity of the developed approach was that structures with disorder of different types were used at the training/validation and inference stages. 

The aim of the current contribution is to estimate the instability of a screening pipeline and evaluate robustness of different backbones of it.
The main goal is evaluating impact of the less relevant to the target domain data impact on the results.
Using the data and the approach presented in \cite{eremin2022hybrid}, we have generated 13 subdatasets based on the disordering level and thus have modeled the changes in the training data. 
It is shown that the Random Forest algorithm \cite{breiman2001random} is robust and stable against the training data changes, while the Allegro \cite{allegro} neural network is not.
Its predictions could become completely different; the assessment was done not only in a quantitative root squared error (RMSE) way, but as well in a qualitative manner. We have also demonstrated that the pre-training on the corresponding Aflow dataset \cite{CURTAROLO2012227} slice increases stability and introduced an elementary sequential training trick to increase algorithm robustness without involving additional data.

\section{Background} \label{background}

\subsection{Doping and screening}
For many years, modeling of doping and chemical modification of known compounds has attracted researchers from both practical and theoretical points of view. 
Practical interest in doped or chemically modified systems is certainly associated with the possibilities of targeted modification of various properties of existing functional materials, diversification of the technology stack, and disruptive innovations. 
In theoretical research, the approaches based on chemical substitutions can potentially drastically enrich the chemistry of known structure types. \cite{merchant2023scaling}
The use of one-to-many chemical modifications also served to create computational datasets, such as the AFLOWLIB collection \cite{curtarolo2012aflow}, some entries of which are crystalline structures that exist only as a computer model. 
In this way, the use of data-driven approaches can be used both for intelligent sampling of chemical search spaces \cite{yuan2023active} and for accelerating screening over the complex compositional-configuration spaces \cite{merchant2023scaling, eremin2024perovskites}.

For many complex crystal structures, the study of the thermodynamics of disordering in crystalline structures can be important in interpreting experimental data of structural research.\cite{solokha2020new}
Moreover, the idea of the thermodynamic favorability of certain types of defects can be used to significantly simplify target search spaces during the search for potentially new compositions.\cite{eremin2022hybrid}

Despite the recently challenges in the data-inspired discoveries of new materials spotlighted recently\cite{leeman2024challenges}, the perspectives of high-throughput approaches development in chemistry and materials science are obvious.
We believe that in the field of natural sciences, parity will be maintained between interpretable and non-interpretable, but probably more transferable, models particularly for data-efficient and narrower targeted applications.
For this reason, we address both branches of them in this work -- classical approaches and neural networks.

\subsection{Classic machine learning models}
In this work, we used Random Forest \cite{breiman2001random} and Gradient Boosting approaches such as XGBoost \cite{xgboost}, LightGBM \cite{LightGBM} and CatBoost \cite{catboost}. Random Forest is a technique in ensemble learning that creates numerous decision trees during training, making it applicable for classification, regression, and various other tasks. In regression, it returns the mean or average prediction made by the individual trees. 
Gradient boosting is a prominent machine learning algorithm utilized for various tasks such as regression and classification. It produces a prediction model by combining multiple weak prediction models, typically simple decision trees. 
It is worth noting that all the aforementioned algorithms take as input a specially prepared table of features, and not a chemical system in its raw form.

\subsection{Allegro neural network}
GNNs are deep learning techniques specifically designed to handle graph data and widely used in many areas from socials networks (\cite{ZHOU202057}) to natural science, including materials science domain (\cite{duval2023hitchhiker}). 
The Allegro architecture \cite{allegro}, which was used for the purpose of this study, is a state-of-the-art strictly local equivariant graph neural network that learns representations related to pairs of neighboring atoms by utilizing two latent spaces. 
The first is an invariant latent space, consisting of scalar features, and the second is an equivariant latent space, capable of processing tensors of any rank. 
These two latent spaces interact with each other at each layer. Finally, a multi-layer perceptron computes the final pairwise energy using the scalar features from the final layer. 
Graph neural networks does not require descriptor based features, work with only spacial information about atomic numbers and coordinates, and account for the periodic boundary conditions of crystals.

\section{Data} 
The considered dataset was presented and fully described in \cite{eremin2022hybrid}; it is available on demand. 
In this study, we used the original dataset without any modifications.

\subsection{Overall description} \label{data: overall description}
The dataset consists of doped Sc-M intermetallic systems, scandium-rich Mackay-type quasicrystal approximants.
The dopants M are four metals: Pt, Pd, Ir and Rh, and thus four corresponding subdatasets were formed.
On the A4 and A5 sites corresponding to the intercluster regions, the original Sc$_{60}$M$_{13}$ crystal structure was disordered by adding vacancies or Sc substitutions by metal M with a step of 2 atoms.
Then the structures with low Sc content (less than 0.667 for Pt and Pd, less than 0.733 for Ir and Rh) were excluded from the further consideration due to the well-known competing phases.
For the rest part of the search space, one random realization of crystal structure with each of the remaining compositions was chosen and computed using Vienna Ab initio Simulation Package (VASP) \cite{PhysRevB.47.558} Density Function Theory (DFT)-solver.

Except for the crystal structures prior to DFT-based relaxation of them, the dataset contains hand-crafted geometrical/topological descriptors (described in \cite{eremin2022hybrid, 10.1063/1.5130082}), which allows using both tabular-based classical models and neural networks independently of each other.

\subsection{The term "limit"}
\begin{figure}[ht]
\centering
\includegraphics[width=1\textwidth]{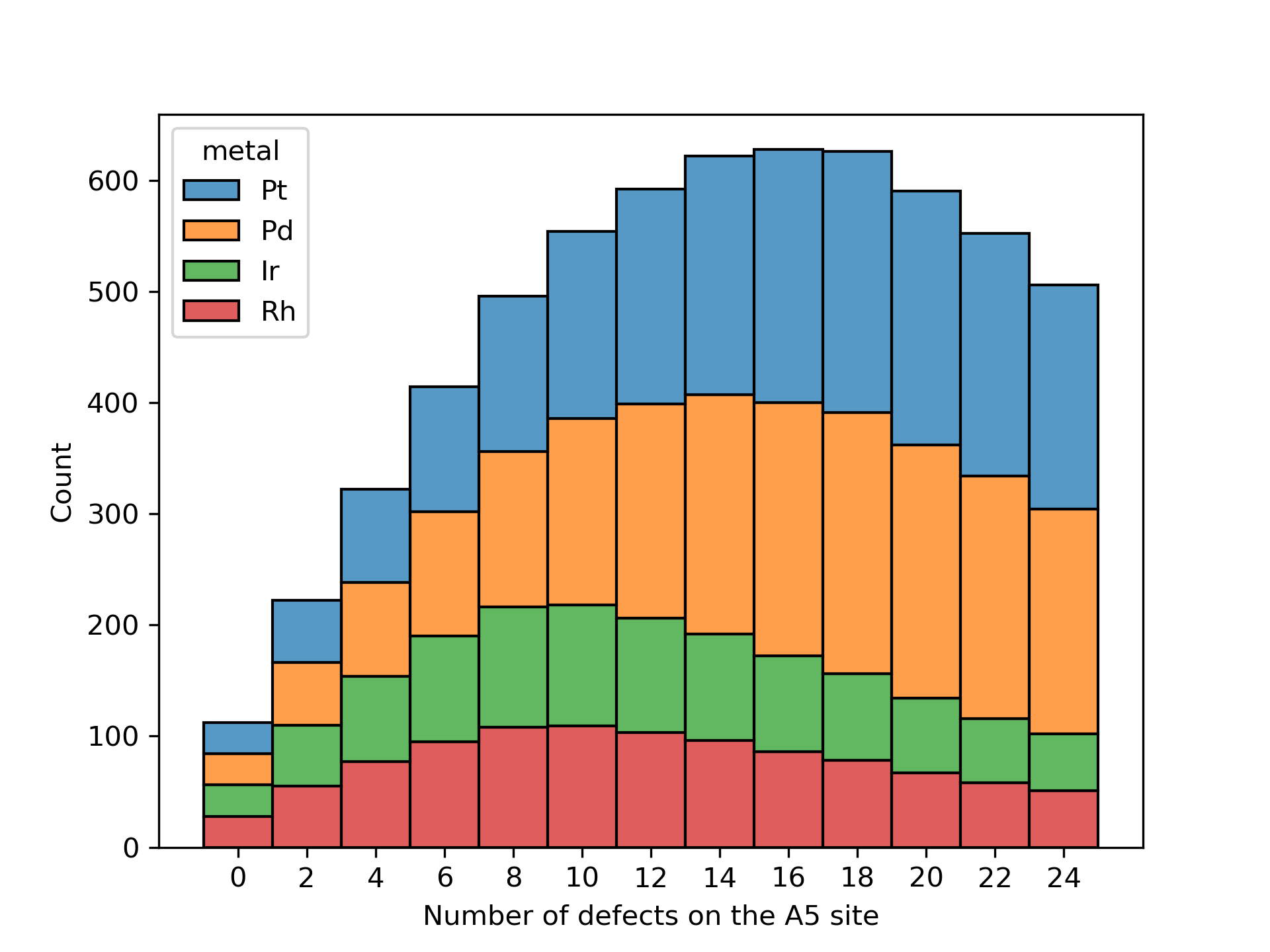}
\caption{Number of systems with a given number of defects on the A5 site of the original crystal structure. The subdataset corresponding to the limit of $x$ comprises all systems with the number of defects on A5 \emph{less or equal} to $x$}
\label{im: limits_size}
\end{figure}

The DFT-calculations showed that defects on the A4 site are more energetically favorable than that on A5 one, and thus we used numbers of defects on A5 as a quantitative measurement of sample similarity. 
So, we introduce the 'limit' term corresponding to the maximum number of defects on the A5 site allowed to add to a certain subdataset. 
The maximum limit in the dataset equals the multiplicity of the A5 site, 24, and the minimum is 0. 
Thus, the entire dataset can be represented as 13 nested datasets, with the limit from 0 to 24. 
The obtained subdataset sizes depending on the limit are shown in Figure \ref{im: limits_size}.

\subsection{Splits} \label{data: splits}
Firstly, for each dopant the inference dataset containing all possible defects on the A4 site with respect to symmetry was created. Despite the inference datasets do not contain target energies, they are precisely the purpose of screening and contain potential new materials, and thus in this work they are used for the final comparison of the quality of the models.

Secondly, each of the four Sc-Pd, Sc-Pt, Sc-Ir, and Sc-Rh datasets is divided into train, validation, and test samples in the same manner: the test set consists only of limit=0 part, while train and val contain all the other systems with the number of defects on A5 from 0 to 24. 
Train/val partition was made once for the full (limit=24) dataset, and then applied accordingly to the subdatasets with limits less than 24.
The precise sizes of all the data samples obtained are represented in Table \ref{tab: sample sizes} 

\begin{table}[h]
    \caption{Precise sizes of all the data samples for each of the studied dopants}
    \begin{center}
        \begin{tabular}{ccccc}
        Composition  & Training set & Validation set & Test set & Total number \\
        \hline
        Sc-Pt, Sc-Pd & 1896         & 211            & 63       & 2170                    \\
        Sc-Rh, Sc-Ir & 909          & 102            & 63       & 1074                   
        \end{tabular}
        \label{tab: sample sizes}
    \end{center}
\end{table}

\subsection{Target energies}
\begin{figure}[ht]
\centering
\includegraphics[width=0.7\textwidth]{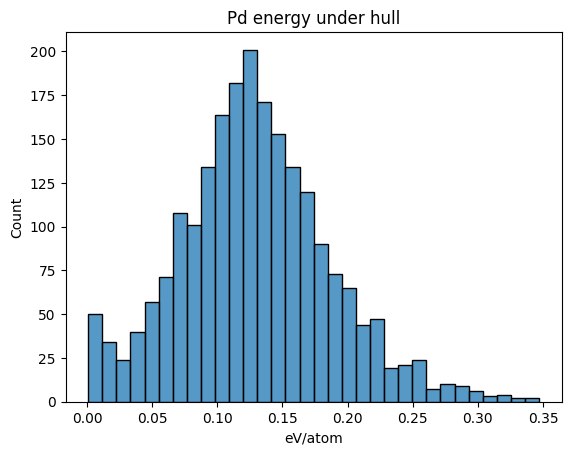}
\caption{Histogram of energies above the hull histogram for the Sc-Pd dataset}
\label{im: energy_size}
\end{figure}

The training part of the entire dataset contains three target properties: energy of above the convex the hull, formation energy, and structure energy in the relaxed state. 
Each of the energies can be calculated from one another deterministically using the following formulas. 
For a certain ${Sc_{x}M_{y}}$ configuration with the characteristic relaxed energy of $E_{Sc_{x}M_{y}}$, the formation energy per atom $E_{formation}$ could be calculated using formula \ref{eq: formation}, where $E_{Sc}$ and $E_{M}$ are the characteristic energies per atom of the Sc and other metal M constituents in a pure metal state, respectively. 
The formation energy $E_{formation}$ transformation into energy above the hull $E_{abovehull}$ can be seen in Figure \ref{eq: ehull}. 
\begin{equation}
\label{eq: formation}
    E_{formation} = \frac{E_{Sc_{x}M_{y}} - xE_{Sc} - yE_{M}}{x+y}
\end{equation}

\begin{equation}
\label{eq: ehull}
    E_{abovehull} = E_{formation} - E_{hull}
\end{equation}
It is worth noting that the energy above the hull is, by design, positive for most systems, and that the ultimate goal of the screening pipeline is to search for crystals that have negative (or at least slightly positive) energy: only such systems meets synthesizability restrictions.
Figure \ref{im: energy_size} illustrates the distribution of the energies above the hull in the Sc-Pd combined dataset.

\section{Methods}
In this work, we considered such classical models as Random Forest (RF),  CatBoost, XGBoost and LightGBM, and the Allegro neural network. 

The energy above the convex hull is to be predicted. For classical algorithms, the hand-made features based on the aforementioned (section \ref{data: overall description}) descriptors were used. 

For each architecture and for each dataset, we used the following procedure: 13 models were trained (one for each limit=$x$ subdataset) and compared on the single test sample. Then an additional comparison was made on the inference dataset

\subsection{Pre-training}
Following common practices of neural network training, we prepared a pretrained model. We trained the Allegro graph neural network \cite{allegro} on the Aflow \cite{CURTAROLO2012227} database slice, consisting of the structures with at least two species out of Sc, Pt, Pd, Ir, and Rh pair in their compositions.

\subsection{sequential training trick}
In this section, we propose a simple yet quite effective solution based on a pre-training approach, but without involving extra data. First, we trained the model on data that is less-relevant, that is, on every piece of data which limit is not equal to 0. After that, the additional training stage was done on 0-limit data.

\subsection{Training}
For classical models, 5-fold cross-validation was used the predictions of which were then averaged, as for the neural network we used a described in \ref{data: splits} piece of data for validation purposes. 
Root-mean-square error (RMSE) \ref{eq: rmse} was used as error function for training and validation steps.
Random seeds were fixed for all experiments.

\begin{equation}
\label{eq: rmse} 
RMSE(y, \hat{y})= \sqrt{\frac{\sum_{i=0}^{N - 1} (y_i - \hat{y}_i)^2}{N}},
\end{equation}
where $y$ is the model prediction, $\hat{y}$ is the target and the $N$ is the dataset size.

\subsection{Test}
Following common practices, we compare model performances on the dataset described in \ref{data: splits} by using the RMSE \ref{eq: rmse} scores of their predictions.
Nevertheless, RMSE itself is not informative enough for screening tasks.
For the purpose of this study, we introduce the custom metric referred to as SimilarityScore \ref{eq: custom_mape}, which allows us to evaluate the similarity of predictions of the models trained on different limit=$x$ subdatasets limits adjusted for target mean deviation. 

\begin{equation}
\label{eq: custom_mape}
    SimilarityScore_{i,j} = 1 - \frac{|RMSE_{i} - RMSE_{j}|}{\overline{y}},
\end{equation}
where $RMSE_{i}$ is the test RMSE score of the model trained on limit=$i$ subdataset and the $\overline{y}$ is the test dataset mean.

As the prediction for classical ML models, we used an average prediction.
As the prediction for classical ML models, we use an average of model predictions from the cross-validation loop. One pipeline was assembled using the scikit-learn API for all models, in which they are alternately changed.

\subsection{Inference} \label{methods: inference}
Being focused on finding new materials while compare different approaches, we need to see qualitative, rather than quantitative change in the results of the model. 

The usual purpose of researches in this field described in section \ref{background} is looking for systems whose energy above the hull is less than zero, so the key idea is to switch from regression to binary classification task. This approach allows to qualitatively compare the model trained on a different limit$=x$ subdatasets.

We considered two ways of comparison:
\begin{enumerate}
    \item using  results of the limit=24 model as a ground truth predictions. 
    \item an observation of the percentages and the number of samples that are below the convex hull intersecting between the limits to achieve a symmetrical comparison
\end{enumerate}

\section{Results}
In this section, we compare performance scores and results of classic machine learning models, such as Gradient Boosting and Random Forest, and Allegro model, qualitatively and quantitatively analyze them, demonstrate the consequences of model instability and consider how to handle these.

\subsection{Test}
\begin{figure}[h]
\centering
\includegraphics[width=1\textwidth]{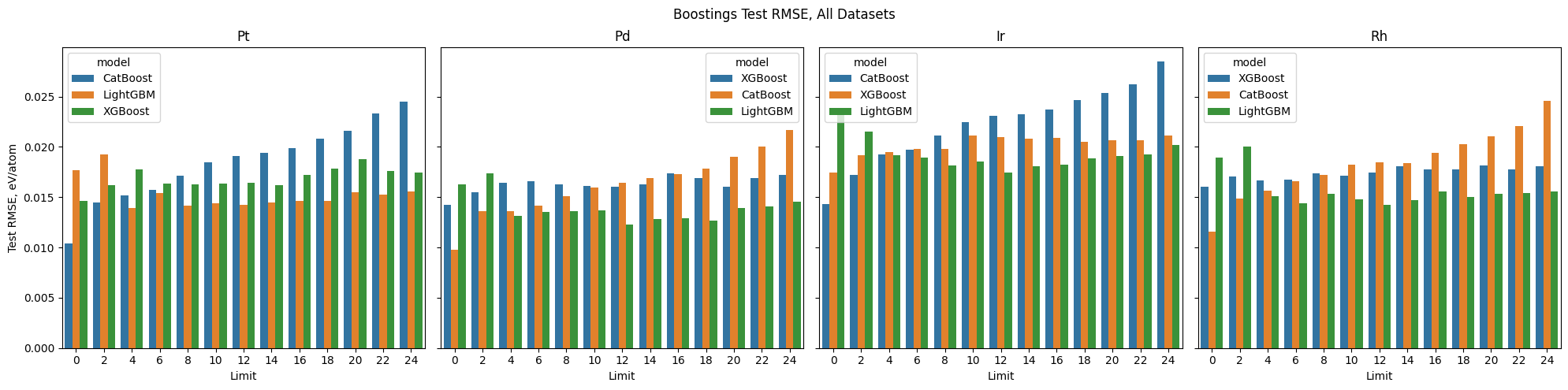}
\caption{Test RMSEs of the gradient boosting models}
\label{im: test_rmse_boosting}
\end{figure}

\begin{figure}[h!]
\centering
\includegraphics[width=0.65\textwidth]{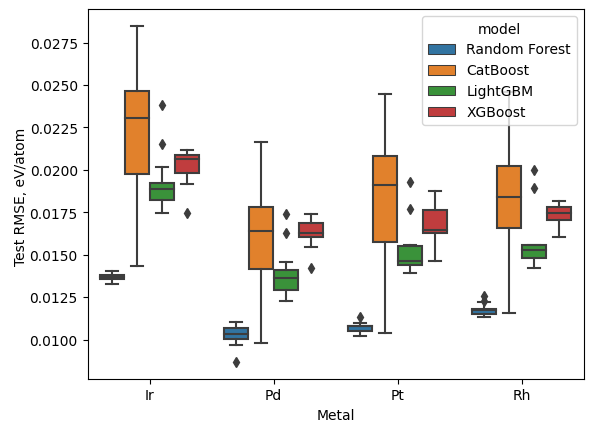}
\caption{Box plots of the obtained test RMSEs for different classical models on different limit=$x$ subdatasets for each dopant, eV/atom}
\label{im: classical}
\end{figure}

Even though RMSEs of all classical Gradient Boosting runs (see Figure \ref{im: test_rmse_boosting}) does not significantly vary (XGBoost) or have a clearly visible pattern (CatBoost and LightGBM), 
all the chosen Gradient Boosting algorithms perform worse than Random Forest approach, which can be seen from Figure \ref{im: classical}

Random Forest have much lower test RMSE compared to that of the tested Gradient Boosting models. 
Random Forest shows test RMSE of \textit{ca}. 0.01 eV/atom, and Gradient Boosting models' scores are of \textit{ca} 0.02 eV/atom.
The latter have a tendency to overfit on the subdatasets with bigger limits (while test sample corresponds to limit=0) and their predictions are highly biased.
Moreover, we checked made and checked the predictions of the trained Gradient Boosting models for the inference set.
The results obtained were rather unrealistic and inconsistent.
For this reason, Gradient Boostings were excluded from further consideration and only the Random Forest model is used. 

\begin{figure}[h]
\centering
\includegraphics[width=0.8\textwidth]{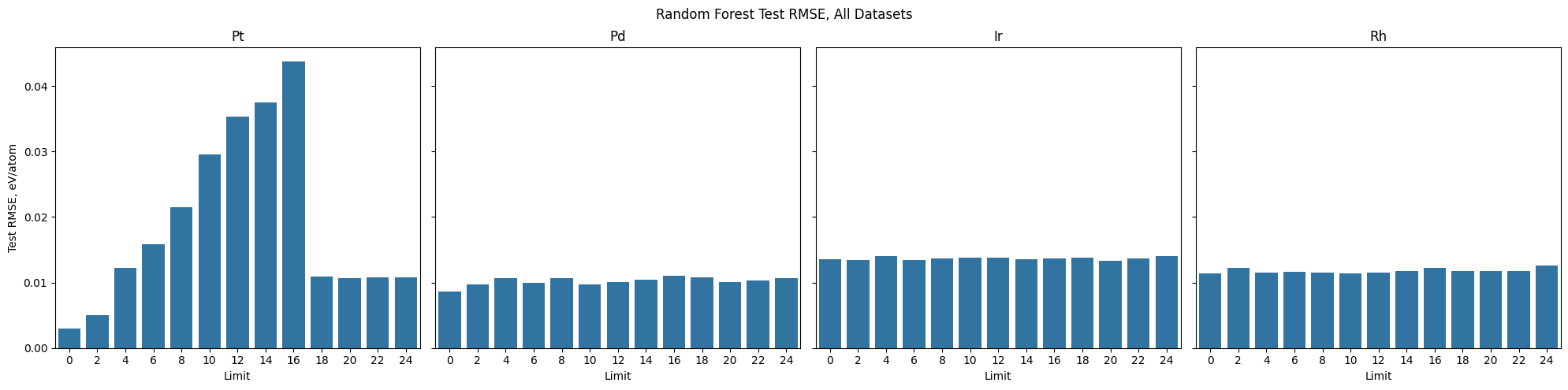}
\caption{Random Forest Test RMSE, eV/atom}
\label{im: test_rmse_rf}
\end{figure}

Random Forest provides stable results, i.e., the test RMSEs are found to fluctuate within a few meV/atom for three out of four datasets (except for the Sc-Pd systems) as shown in Figure \ref{im: test_rmse_rf}. 

\begin{figure}[h]
\centering
\includegraphics[width=1\textwidth]{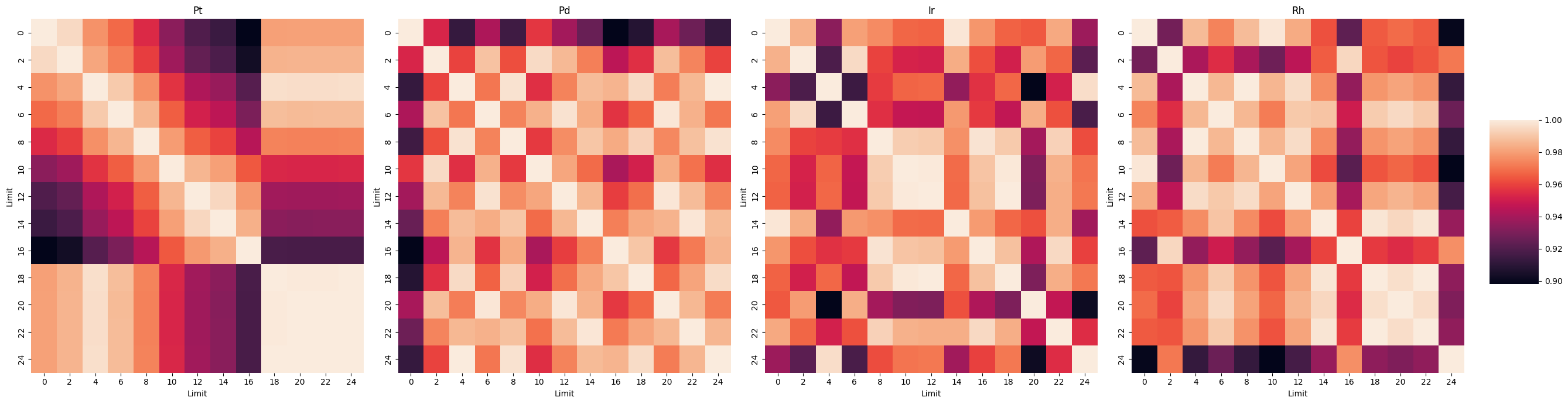}
\caption{Random Forest pair-wise SimilarityScore on the test sample}
\label{im: test_ss_rf}
\end{figure}

In Figure \ref{im: test_ss_rf}, the SimilarityScores matrices also illustrate the Random Forest stability: sharp changes are common only for the Sc-Pt systems.

\begin{figure}[h!]
\centering
\includegraphics[width=0.57\textwidth]{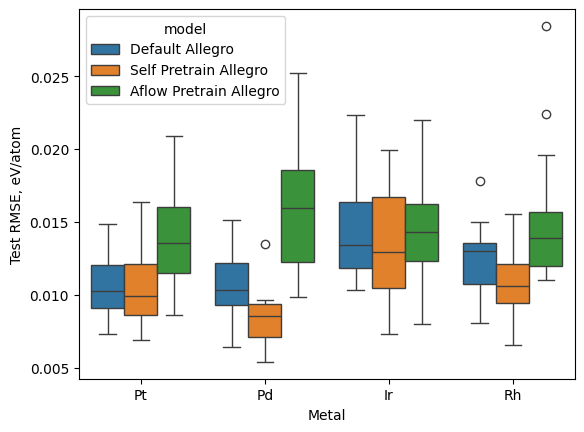}
\caption{Box plots of the obtained test RMSEs for different Allegro training approaches on different limit=$x$ subdatasets for each dopant, eV/atom} 
\label{im: allegro_full_test}
\end{figure}

Different neural network training approaches perform about the same RMSE ranges of 0.010–0.015 eV/atom on the test set. 
As it is shown in Figure \ref{im: allegro_full_test}, the self-pre-trained Allegro model's errors are always smaller than that of the ordinarily trained one. 
The Aflow-pretrained version, in turn, shows bigger errors. 

\begin{figure}[h]
\centering
\includegraphics[width=1\textwidth]{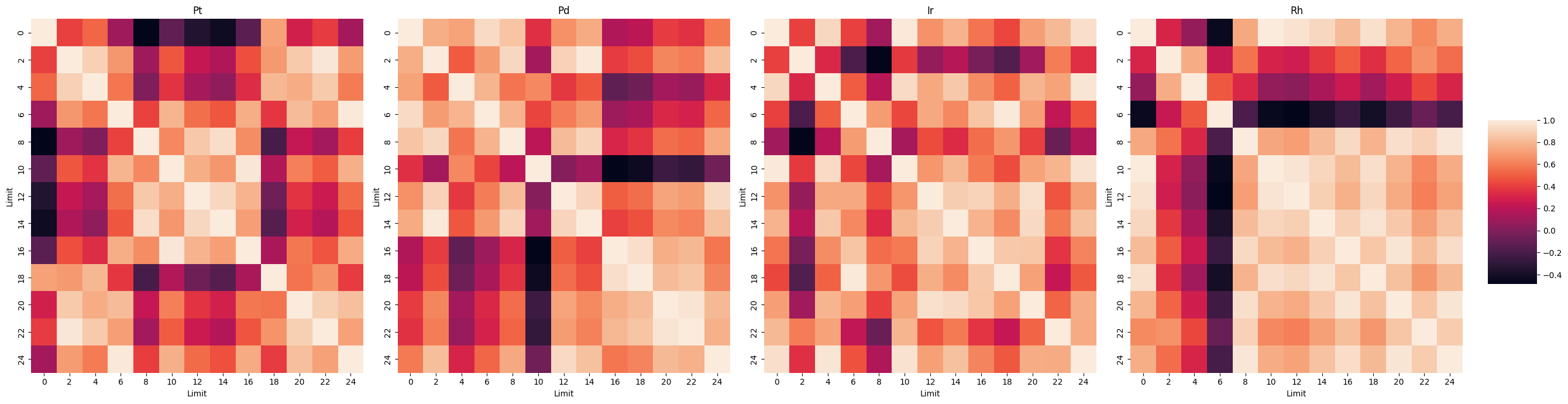}
\caption{Heatmaps of the SimilarityScore obtained using the Aflow-pretrained Allegro model on the test sample}
\label{im: aflow_metric}
\end{figure}

\begin{figure}[h]
\centering
\includegraphics[width=1\textwidth]{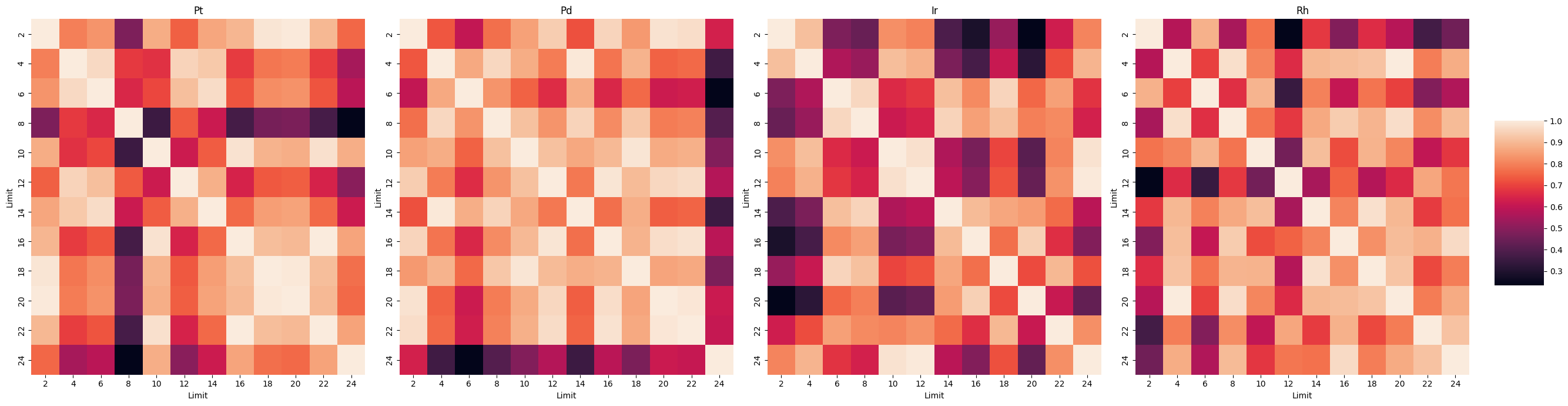}
\caption{Heatmaps of the SimilarityScore obtained using the self-pretrained Allegro model on the test sample}
\label{im: pretrain_metric}
\end{figure}

\begin{figure}[h]
\centering
\includegraphics[width=1\textwidth]{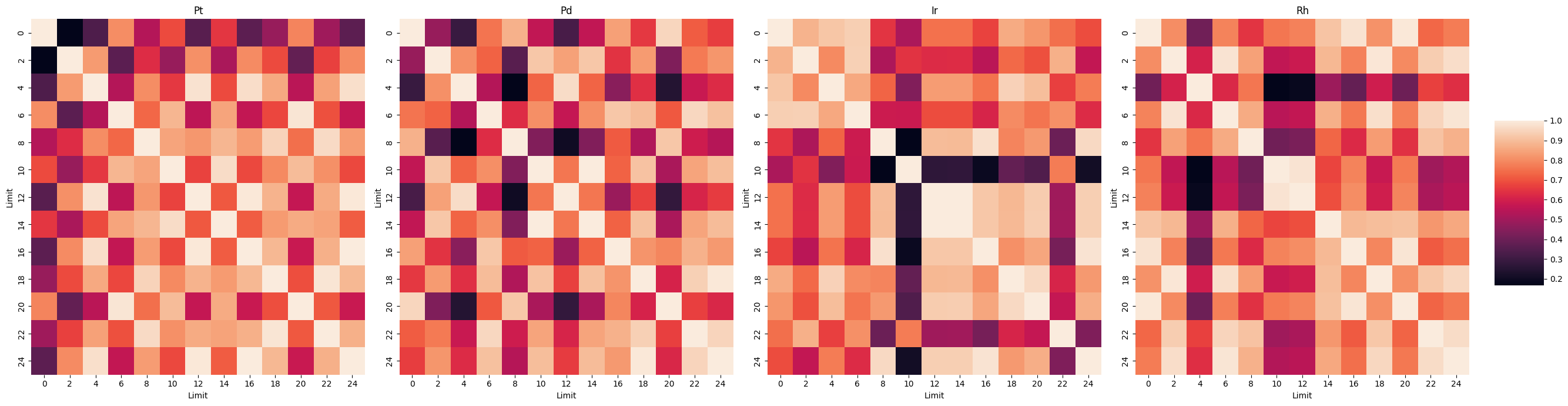}
 \caption{Heatmaps of the SimilarityScore obtained using the Allegro model without pretraining on the test sample}
\label{im: base_metric}
\end{figure}

In contrast to the similar RMSE values of the models tested, the Similarity Score shows pronounce differences between the model predictions at a qualitative level.
The Aflow- (Figure \ref{im: aflow_metric}) and self-pretrained (Figure \ref{im: pretrain_metric}) Allegro models provide heatmaps that illustrates larger Similarity Scores than that on the ordinarily trained network (Figure \ref{im: base_metric}). 

Moreover, for the self-pretrained Allegro the light areas are much more clearly located. If a light bar appears after a certain limit, it means that it is likely to continue at the next limits. For some metals, there is a pattern where the model gets good results up to a certain limit, comprising the relevant data only. 
Then, the model shows a worse result due to getting a little less relevant data. 
However, the model being provided by a large amount of both the relevant (limit=0) and less relevant (limit=$x$, $x>0$) data shows better results again.
It can be assumed that by perceiving the overall data, it increases the robustness of predictions.

\subsection{Inference}

\begin{figure}[h]
\begin{minipage}[ht]{0.49\linewidth}
\center{\includegraphics[width=0.9\linewidth]{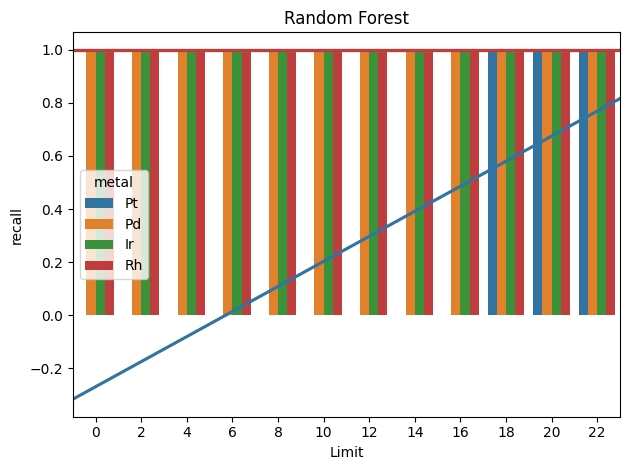} \\ a)}
\end{minipage}
\hfill
\begin{minipage}[ht]{0.49\linewidth}
\center{\includegraphics[width=0.9\linewidth]{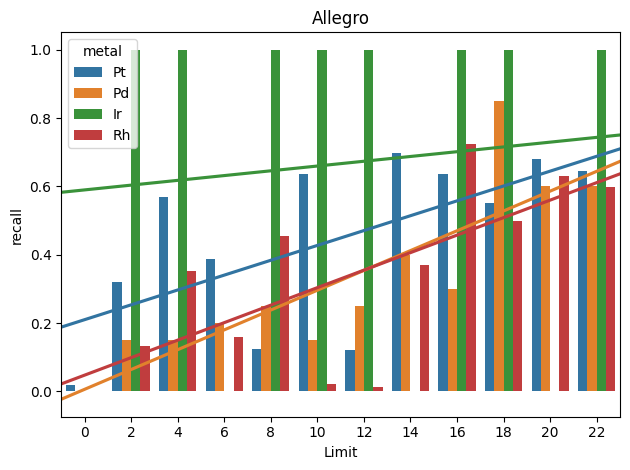} \\ b)}
\end{minipage}
\caption{(a) Random Forest test recall, considering limit=24 model prediction as the ground truth (b) Allegro test recall, considering limit=24 model prediction as the ground truth}
\label{im: recalls}
\end{figure}

The comparison between the limit=24 models and all the others shows that the Random Forest is robust, and the Allegro model is not.
The recall metric computed considering the limit=24 predictions as the ground truth is always 1 for Random Forests (see Figure \ref{im: recalls} a) and vary from 0 to 1 for Allegro (see Figure \ref{im: recalls} b).



\begin{figure}[h]
\centering
\includegraphics[width=1\textwidth]{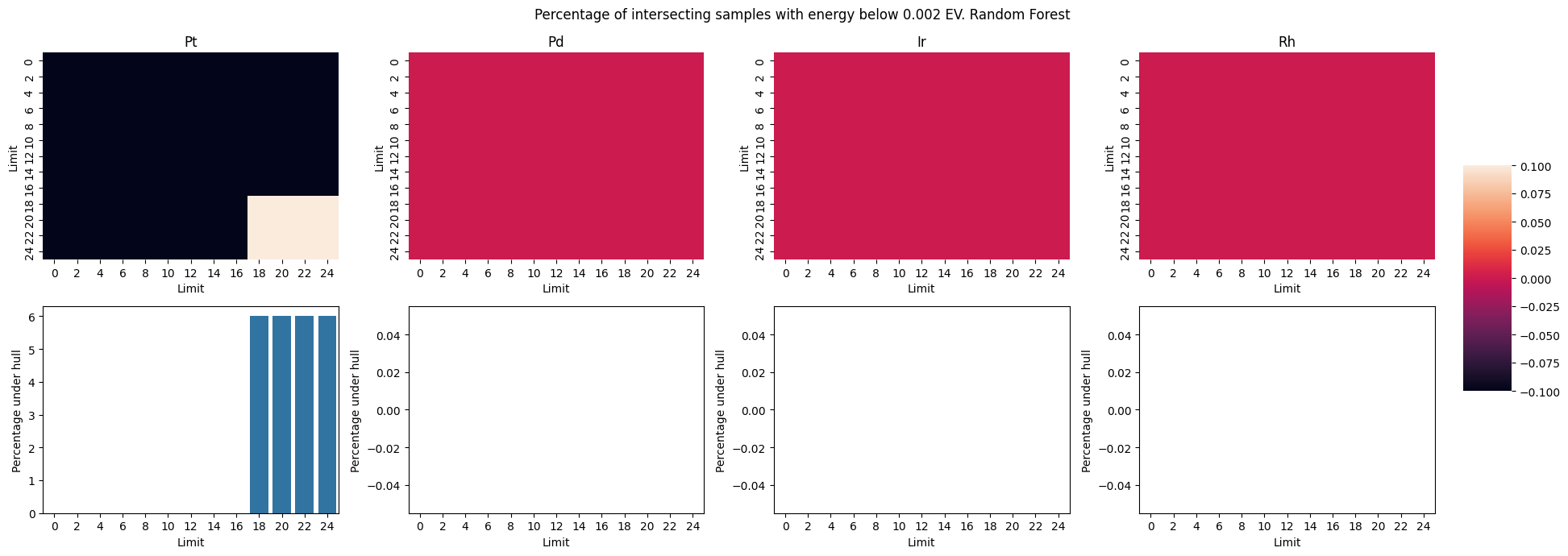}
\caption{Random Forest Inference}
\label{im: rf inference}
\end{figure}

The pairwise comparison of limits described in \ref{methods: inference} highlights the apparent difference between the models, trained on different subdatasets with limit=$x$. 
Random Forest behaves stable, showing not the chaotic changes, but a visible transition (see Figure \ref{im: rf inference}) from different to similar runs. The empty graphs and solid-colored heatmaps in Figure \ref{im: rf inference} show that there are no structures predicted below the convex hull. 

\begin{figure}[hb]
\centering
\includegraphics[width=1\textwidth]{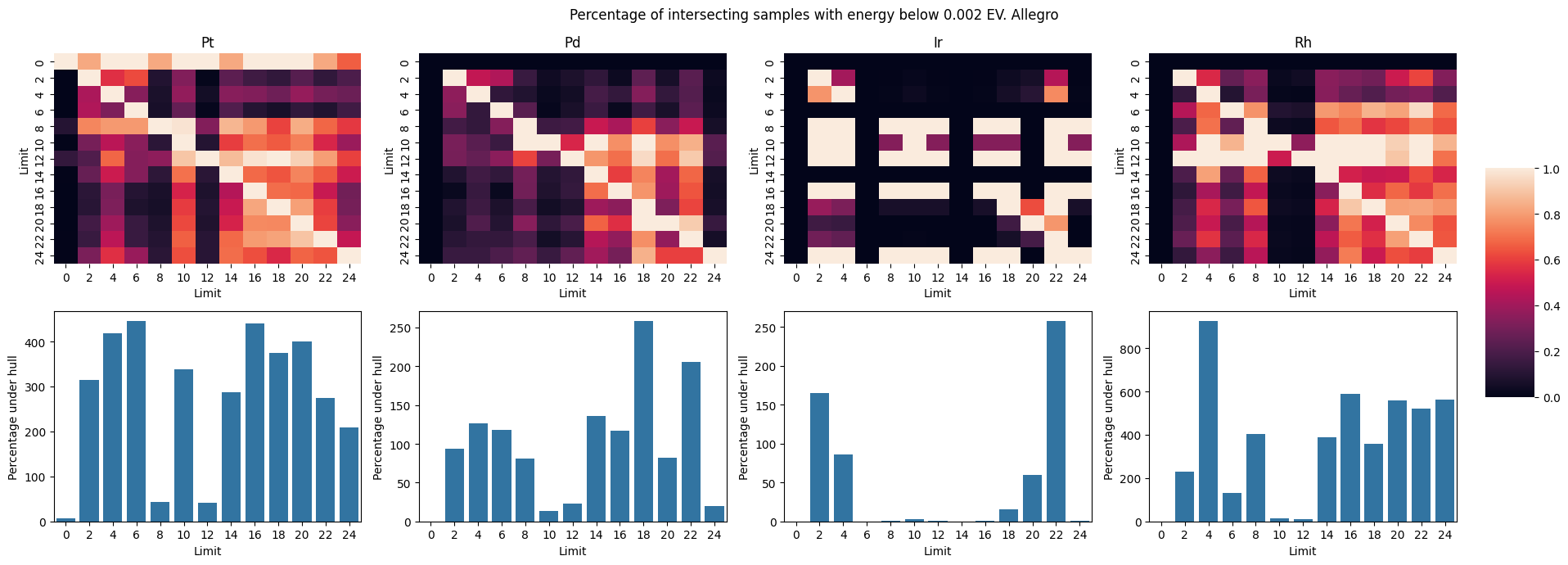}
\caption{Allegro Inference}
\label{im: allegro inference}
\end{figure}

\begin{figure}[h]
\centering
\includegraphics[width=1\textwidth]{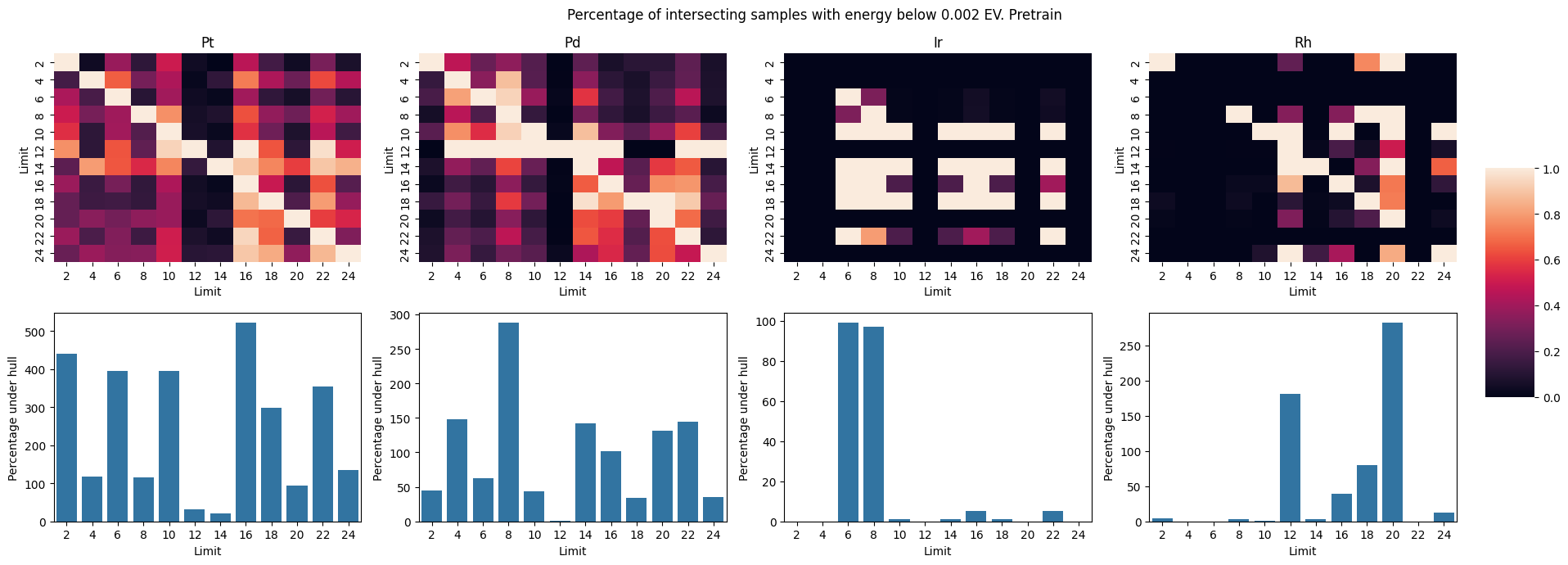}
\caption{Allegro Self-pretrained Inference}
\label{im: allegro pretrain inference}
\end{figure}

\begin{figure}[h]
\centering
\includegraphics[width=1\textwidth]{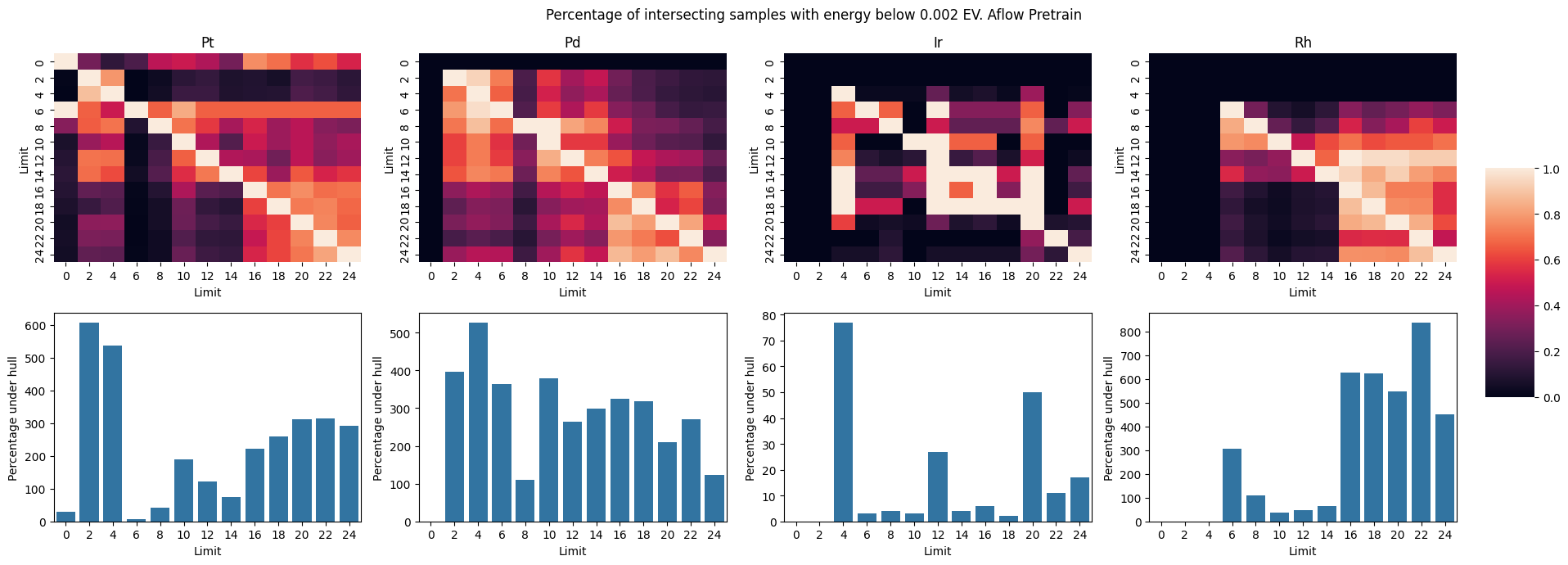}
\caption{Allegro Aflow Pretrained Inference}
\label{im: allegro aflow pretrain inference}
\end{figure}

Moreover, there are similarities between the results shown in inference heatmaps (see Figures \ref{im: allegro inference}, \ref{im: allegro pretrain inference}, \ref{im: allegro aflow pretrain inference}) and test heatmaps (see Figures \ref{im: aflow_metric}, \ref{im: pretrain_metric}, \ref{im: base_metric}) 

The heatmaps indicate the instability of neural networks, as this behavior is not observed in the results of the Random Forest. Despite the fact that the results may vary, they are not random, and there is some kind of dependence factor. As it can be seen in Figure \ref{im: allegro inference}, there is a trade-off between the relevance of certain data points and their amounts provided to the model. 
In the middle range of the introduced limits, a small change in the limit leads to large changes in the results. 
This observation clearly illustrates the following principle -- training data should be either small and strictly relevant or big and less relevant. 
From the dependencies obtained for Random Forest, this can also be seen.

As shown in Figures \ref{im: allegro pretrain inference}, \ref{im: allegro aflow pretrain inference}, the trade-off is definitely more visible and after a certain limit on each metal there is a great similarity between structures that are predicted below the convex hull. 
This also shows that there is a certain “relevance threshold” for palladium and platinum. 
With pretrained network, we can overcome unstable behavior of neural-network-based approaches without generation of additional data. 
Thresholds differ from those seen in the pictures of sequential training and regular training. 
This is most likely due to the fact that the model was already familiar with similar data patterns, and its ordinary training requires less data particular for a certain task. 
For rhodium, the threshold is also apparently observed, which shows that when training on data with a limit of 6 or more, the model predicts structures below the hull. 
The results for iridium do not allow us to unambiguously assess the effectiveness of the method proposed.


\section{Conclusion}
    ML-based predictors of thermodynamic properties of crystals are widely used in materials science, but the final goal always lies beyond the regression task.
    Thus, in conducting this study, the primary focus was on the classification task within the domain, emphasizing the need to maintain a clear perspective on this objective throughout the analysis. While the RMSE metric is commonly used to evaluate model performance, it was observed that RMSE alone does not provide a comprehensive view of the model performance. 
    Our study reveals that even small changes in training data and RMSE on a test set can lead to significant variations on the inference set, particularly evident in neural networks trained on small datasets. This lack of robustness in neural-network-based solutions when faced with diverse datasets was highlighted, prompting the use of Random Forest models to corroborate and further validate the results obtained. The Random Forest models, although offering stability, also have limitations, especially when resting upon descriptors that may pose challenges in data collection, or when systems represented by these descriptors are indistinguishable from each other. 
    
    Moreover, the essential consideration highlighted in the study was the trade-off between the relevance and quantity of training data, indicating that simply having more data does not always equate to an improved model performance. A simple yet effective approach suggested was to self-pretrain models on less relevant data before fine-tuning them on more specific parts of the original dataset, to enhance their efficacy in handling diverse data sources and improving overall performance.

\bibliographystyle{unsrtnat}
\bibliography{main}

\begin{thebibliography}{32}
\providecommand{\natexlab}[1]{#1}
\providecommand{\url}[1]{\texttt{#1}}
\expandafter\ifx\csname urlstyle\endcsname\relax
  \providecommand{\doi}[1]{doi: #1}\else
  \providecommand{\doi}{doi: \begingroup \urlstyle{rm}\Url}\fi

\bibitem[Roser(2022)]{owid-brief-history-of-ai}
Max Roser.
\newblock The brief history of artificial intelligence: the world has changed
  fast — what might be next?
\newblock \emph{Our World in Data}, 2022.
\newblock https://ourworldindata.org/brief-history-of-ai.

\bibitem[Zhou et~al.(2023)Zhou, Li, Li, Yu, Liu, Wang, Zhang, Ji, Yan, He,
  et~al.]{zhou2023comprehensive}
Ce~Zhou, Qian Li, Chen Li, Jun Yu, Yixin Liu, Guangjing Wang, Kai Zhang, Cheng
  Ji, Qiben Yan, Lifang He, et~al.
\newblock A comprehensive survey on pretrained foundation models: A history
  from bert to chatgpt.
\newblock \emph{arXiv preprint arXiv:2302.09419}, 2023.

\bibitem[Haenlein and Kaplan(2019)]{haenlein2019brief}
Michael Haenlein and Andreas Kaplan.
\newblock A brief history of artificial intelligence: On the past, present, and
  future of artificial intelligence.
\newblock \emph{California management review}, 61\penalty0 (4):\penalty0 5--14,
  2019.

\bibitem[Han et~al.(2023)Han, Yoon, Kim, Lee, and Lee]{han2023revolutionizing}
Ri~Han, Hongryul Yoon, Gahee Kim, Hyundo Lee, and Yoonji Lee.
\newblock Revolutionizing medicinal chemistry: the application of artificial
  intelligence (ai) in early drug discovery.
\newblock \emph{Pharmaceuticals}, 16\penalty0 (9):\penalty0 1259, 2023.

\bibitem[Ivanenkov et~al.(2023)Ivanenkov, Zagribelnyy, Malyshev, Evteev,
  Terentiev, Kamya, Bezrukov, Aliper, Ren, and
  Zhavoronkov]{ivanenkov2023hitchhiker}
Yan Ivanenkov, Bogdan Zagribelnyy, Alex Malyshev, Sergei Evteev, Victor
  Terentiev, Petrina Kamya, Dmitry Bezrukov, Alex Aliper, Feng Ren, and Alex
  Zhavoronkov.
\newblock The hitchhiker’s guide to deep learning driven generative
  chemistry.
\newblock \emph{ACS Medicinal Chemistry Letters}, 14\penalty0 (7):\penalty0
  901--915, 2023.

\bibitem[Alwosheel et~al.(2021)Alwosheel, van Cranenburgh, and
  Chorus]{alwosheel2021did}
Ahmad Alwosheel, Sander van Cranenburgh, and Caspar~G Chorus.
\newblock Why did you predict that? towards explainable artificial neural
  networks for travel demand analysis.
\newblock \emph{Transportation Research Part C: Emerging Technologies},
  128:\penalty0 103143, 2021.

\bibitem[Hamon et~al.(2020)Hamon, Junklewitz, Sanchez,
  et~al.]{hamon2020robustness}
Ronan Hamon, Henrik Junklewitz, Ignacio Sanchez, et~al.
\newblock Robustness and explainability of artificial intelligence.
\newblock \emph{Publications Office of the European Union}, 207:\penalty0 2020,
  2020.

\bibitem[Kejriwal et~al.(2024)Kejriwal, Kildebeck, Steininger, and
  Shrivastava]{kejriwal2024challengersjoshoconnorzendaya}
Mayank Kejriwal, Eric Kildebeck, Robert Steininger, and Abhinav Shrivastava.
\newblock Challenges, evaluation and opportunities for open-world learning.
\newblock \emph{Nature Machine Intelligence}, pages 1--9, 2024.

\bibitem[Goyal et~al.(2023)Goyal, Doddapaneni, Khapra, and
  Ravindran]{10.1145/3593042}
Shreya Goyal, Sumanth Doddapaneni, Mitesh~M. Khapra, and Balaraman Ravindran.
\newblock A survey of adversarial defenses and robustness in nlp.
\newblock \emph{ACM Comput. Surv.}, 55\penalty0 (14s), jul 2023.
\newblock ISSN 0360-0300.
\newblock \doi{10.1145/3593042}.
\newblock URL \url{https://doi.org/10.1145/3593042}.

\bibitem[Wang et~al.(2021)Wang, Wang, and Yang]{wang2021measure}
Xuezhi Wang, Haohan Wang, and Diyi Yang.
\newblock Measure and improve robustness in nlp models: A survey.
\newblock \emph{arXiv preprint arXiv:2112.08313}, 2021.

\bibitem[Liu and Jin(2023)]{LIU2023175}
Jia Liu and Yaochu Jin.
\newblock A comprehensive survey of robust deep learning in computer vision.
\newblock \emph{Journal of Automation and Intelligence}, 2\penalty0
  (4):\penalty0 175--195, 2023.
\newblock ISSN 2949-8554.
\newblock \doi{https://doi.org/10.1016/j.jai.2023.10.002}.
\newblock URL
  \url{https://www.sciencedirect.com/science/article/pii/S294985542300045X}.

\bibitem[Li et~al.(2024)Li, Xie, Guo, Yang, and Xiao]{li2024survey}
Yanjie Li, Bin Xie, Songtao Guo, Yuanyuan Yang, and Bin Xiao.
\newblock A survey of robustness and safety of 2d and 3d deep learning models
  against adversarial attacks.
\newblock \emph{ACM Computing Surveys}, 56\penalty0 (6):\penalty0 1--37, 2024.

\bibitem[Merchant et~al.(2023)Merchant, Batzner, Schoenholz, Aykol, Cheon, and
  Cubuk]{merchant2023scaling}
Amil Merchant, Simon Batzner, Samuel~S Schoenholz, Muratahan Aykol, Gowoon
  Cheon, and Ekin~Dogus Cubuk.
\newblock Scaling deep learning for materials discovery.
\newblock \emph{Nature}, pages 1--6, 2023.

\bibitem[Chen et~al.(2024)Chen, Nguyen, Lee, Baker, Karakoti, Lauw, Owen,
  Mueller, Bilodeau, Murugesan, and Troyer]{Chen2024AcceleratingCM}
Chi Chen, Dan~Thien Nguyen, Shannon~J. Lee, Nathan~A. Baker, Ajay~S. Karakoti,
  Linda Lauw, Craig Owen, Karl~T. Mueller, Brian~A. Bilodeau, Vijayakumar
  Murugesan, and Matthias Troyer.
\newblock Accelerating computational materials discovery with artificial
  intelligence and cloud high-performance computing: from large-scale screening
  to experimental validation.
\newblock 2024.
\newblock URL \url{https://api.semanticscholar.org/CorpusID:266843938}.

\bibitem[Jain et~al.(2013)Jain, Ong, Hautier, Chen, Richards, Dacek, Cholia,
  Gunter, Skinner, Ceder, et~al.]{jain2013commentary}
Anubhav Jain, Shyue~Ping Ong, Geoffroy Hautier, Wei Chen, William~Davidson
  Richards, Stephen Dacek, Shreyas Cholia, Dan Gunter, David Skinner, Gerbrand
  Ceder, et~al.
\newblock Commentary: The materials project: A materials genome approach to
  accelerating materials innovation.
\newblock \emph{APL materials}, 1\penalty0 (1), 2013.

\bibitem[Curtarolo et~al.(2012{\natexlab{a}})Curtarolo, Setyawan, Wang, Xue,
  Yang, Taylor, Nelson, Hart, Sanvito, Buongiorno-Nardelli, Mingo, and
  Levy]{CURTAROLO2012227}
Stefano Curtarolo, Wahyu Setyawan, Shidong Wang, Junkai Xue, Kesong Yang,
  Richard~H. Taylor, Lance~J. Nelson, Gus~L.W. Hart, Stefano Sanvito, Marco
  Buongiorno-Nardelli, Natalio Mingo, and Ohad Levy.
\newblock Aflowlib.org: A distributed materials properties repository from
  high-throughput ab initio calculations.
\newblock \emph{Computational Materials Science}, 58:\penalty0 227--235,
  2012{\natexlab{a}}.
\newblock ISSN 0927-0256.
\newblock \doi{https://doi.org/10.1016/j.commatsci.2012.02.002}.
\newblock URL
  \url{https://www.sciencedirect.com/science/article/pii/S0927025612000687}.

\bibitem[Yuan et~al.(2023)Yuan, Zhou, Peng, Yang, Li, and Wen]{yuan2023active}
Xiaoze Yuan, Yuwei Zhou, Qing Peng, Yong Yang, Yongwang Li, and Xiaodong Wen.
\newblock Active learning to overcome exponential-wall problem for effective
  structure prediction of chemical-disordered materials.
\newblock \emph{npj Computational Materials}, 9\penalty0 (1):\penalty0 12,
  2023.

\bibitem[Huang et~al.(2023)Huang, Lukin, Faleev, Kazeev, Al-Maeeni, Andreeva,
  Ustyuzhanin, Tormasov, Castro~Neto, and Novoselov]{huang2023unveiling}
Pengru Huang, Ruslan Lukin, Maxim Faleev, Nikita Kazeev, Abdalaziz~Rashid
  Al-Maeeni, Daria~V Andreeva, Andrey Ustyuzhanin, Alexander Tormasov,
  AH~Castro~Neto, and Kostya~S Novoselov.
\newblock Unveiling the complex structure-property correlation of defects in 2d
  materials based on high throughput datasets.
\newblock \emph{npj 2D Materials and Applications}, 7\penalty0 (1):\penalty0 6,
  2023.

\bibitem[Eremin et~al.(2024)Eremin, Humonen, Kazakov, Lazarev, Pushkarev, and
  Budennyy]{eremin2024perovskites}
Roman~A. Eremin, Innokentiy~S. Humonen, Alexey~A. Kazakov, Vladimir~D. Lazarev,
  Anatoly~P. Pushkarev, and Semen~A. Budennyy.
\newblock Graph neural networks for predicting structural stability of {Cd}-
  and {Zn}-doped $\gamma$-{CsPbI}$_3$.
\newblock \emph{Computational Materials Science}, 232:\penalty0 112672, 2024.
\newblock ISSN 0927-0256.
\newblock \doi{https://doi.org/10.1016/j.commatsci.2023.112672}.
\newblock URL
  \url{https://www.sciencedirect.com/science/article/pii/S0927025623006663}.

\bibitem[Eremin et~al.(2022)Eremin, Humonen, Zolotarev, Medrish, Zhukov, and
  Budennyy]{eremin2022hybrid}
Roman~A. Eremin, Innokentiy~S. Humonen, Pavel~N. Zolotarev, Inna~V. Medrish,
  Leonid~E. Zhukov, and Semen~A. Budennyy.
\newblock Hybrid {DFT}/data-driven approach for searching for new quasicrystal
  approximants in {Sc-X (X = Rh, Pd, Ir, Pt)} systems.
\newblock \emph{Crystal Growth \& Design}, 22\penalty0 (7):\penalty0
  4570--4581, 2022.
\newblock \doi{doi: 10.1021/acs.cgd.2c00463}.
\newblock URL \url{https://pubs.acs.org/doi/10.1021/acs.cgd.2c00463}.

\bibitem[Breiman(2001)]{breiman2001random}
Leo Breiman.
\newblock Random forests.
\newblock \emph{Machine learning}, 45:\penalty0 5--32, 2001.

\bibitem[Musaelian et~al.(2023)Musaelian, Batzner, Johansson, Sun, Owen,
  Kornbluth, and Kozinsky]{allegro}
Albert Musaelian, Simon Batzner, Anders Johansson, Lixin Sun, Cameron~J. Owen,
  Mordechai Kornbluth, and Boris Kozinsky.
\newblock Learning local equivariant representations for large-scale atomistic
  dynamics.
\newblock \emph{Nature Communications}, 14\penalty0 (1):\penalty0 579, February
  2023.
\newblock ISSN 2041-1723.
\newblock \doi{10.1038/s41467-023-36329-y}.
\newblock URL \url{https://doi.org/10.1038/s41467-023-36329-y}.

\bibitem[Curtarolo et~al.(2012{\natexlab{b}})Curtarolo, Setyawan, Hart,
  Jahnatek, Chepulskii, Taylor, Wang, Xue, Yang, Levy,
  et~al.]{curtarolo2012aflow}
Stefano Curtarolo, Wahyu Setyawan, Gus~LW Hart, Michal Jahnatek, Roman~V
  Chepulskii, Richard~H Taylor, Shidong Wang, Junkai Xue, Kesong Yang, Ohad
  Levy, et~al.
\newblock Aflow: An automatic framework for high-throughput materials
  discovery.
\newblock \emph{Computational Materials Science}, 58:\penalty0 218--226,
  2012{\natexlab{b}}.

\bibitem[Solokha et~al.(2020)Solokha, Eremin, Leisegang, Proserpio,
  Akhmetshina, Gurskaya, Saccone, and De~Negri]{solokha2020new}
Pavlo Solokha, Roman~A Eremin, Tilmann Leisegang, Davide~M Proserpio, Tatiana
  Akhmetshina, Albina Gurskaya, Adriana Saccone, and Serena De~Negri.
\newblock New quasicrystal approximant in the sc--pd system: from topological
  data mining to the bench.
\newblock \emph{Chemistry of Materials}, 32\penalty0 (3):\penalty0 1064--1079,
  2020.

\bibitem[Leeman et~al.(2024)Leeman, Liu, Stiles, Lee, Bhatt, Schoop, and
  Palgrave]{leeman2024challenges}
Josh Leeman, Yuhan Liu, Joseph Stiles, Scott~B Lee, Prajna Bhatt, Leslie~M
  Schoop, and Robert~G Palgrave.
\newblock Challenges in high-throughput inorganic materials prediction and
  autonomous synthesis.
\newblock \emph{PRX Energy}, 3\penalty0 (1):\penalty0 011002, 2024.

\bibitem[Chen and Guestrin(2016)]{xgboost}
Tianqi Chen and Carlos Guestrin.
\newblock Xgboost: A scalable tree boosting system.
\newblock In \emph{Proceedings of the 22nd acm sigkdd international conference
  on knowledge discovery and data mining}, pages 785--794, 2016.

\bibitem[Ke et~al.(2017)Ke, Meng, Finley, Wang, Chen, Ma, Ye, and
  Liu]{LightGBM}
Guolin Ke, Qi~Meng, Thomas Finley, Taifeng Wang, Wei Chen, Weidong Ma, Qiwei
  Ye, and Tie-Yan Liu.
\newblock Lightgbm: A highly efficient gradient boosting decision tree.
\newblock In I.~Guyon, U.~Von Luxburg, S.~Bengio, H.~Wallach, R.~Fergus,
  S.~Vishwanathan, and R.~Garnett, editors, \emph{Advances in Neural
  Information Processing Systems}, volume~30. Curran Associates, Inc., 2017.
\newblock URL
  \url{https://proceedings.neurips.cc/paper_files/paper/2017/file/6449f44a102fde848669bdd9eb6b76fa-Paper.pdf}.

\bibitem[Prokhorenkova et~al.(2018)Prokhorenkova, Gusev, Vorobev, Dorogush, and
  Gulin]{catboost}
Liudmila Prokhorenkova, Gleb Gusev, Aleksandr Vorobev, Anna~Veronika Dorogush,
  and Andrey Gulin.
\newblock Catboost: unbiased boosting with categorical features.
\newblock 2018.

\bibitem[Zhou et~al.(2020)Zhou, Cui, Hu, Zhang, Yang, Liu, Wang, Li, and
  Sun]{ZHOU202057}
Jie Zhou, Ganqu Cui, Shengding Hu, Zhengyan Zhang, Cheng Yang, Zhiyuan Liu,
  Lifeng Wang, Changcheng Li, and Maosong Sun.
\newblock Graph neural networks: A review of methods and applications.
\newblock \emph{AI Open}, 1:\penalty0 57--81, 2020.
\newblock ISSN 2666-6510.
\newblock \doi{https://doi.org/10.1016/j.aiopen.2021.01.001}.
\newblock URL
  \url{https://www.sciencedirect.com/science/article/pii/S2666651021000012}.

\bibitem[Duval et~al.(2023)Duval, Mathis, Joshi, Schmidt, Miret, Malliaros,
  Cohen, Lio, Bengio, and Bronstein]{duval2023hitchhiker}
Alexandre Duval, Simon~V Mathis, Chaitanya~K Joshi, Victor Schmidt, Santiago
  Miret, Fragkiskos~D Malliaros, Taco Cohen, Pietro Lio, Yoshua Bengio, and
  Michael Bronstein.
\newblock A hitchhiker's guide to geometric gnns for 3d atomic systems.
\newblock \emph{arXiv preprint arXiv:2312.07511}, 2023.

\bibitem[Kresse and Hafner(1993)]{PhysRevB.47.558}
G.~Kresse and J.~Hafner.
\newblock Ab initio molecular dynamics for liquid metals.
\newblock \emph{Phys. Rev. B}, 47:\penalty0 558--561, Jan 1993.
\newblock \doi{10.1103/PhysRevB.47.558}.
\newblock URL \url{https://link.aps.org/doi/10.1103/PhysRevB.47.558}.

\bibitem[Eremin et~al.(2019)Eremin, Zolotarev, Leisegang, and
  Solokha]{10.1063/1.5130082}
Roman Eremin, Pavel Zolotarev, Tilmann Leisegang, and Pavlo Solokha.
\newblock {A machine learning approach for predicting formation enthalpy: A
  case study of Mackay-type approximants of icosahedral quasicrystals}.
\newblock \emph{AIP Conference Proceedings}, 2163\penalty0 (1):\penalty0
  020003, 10 2019.
\newblock ISSN 0094-243X.
\newblock \doi{10.1063/1.5130082}.
\newblock URL \url{https://doi.org/10.1063/1.5130082}.

\end{thebibliography}

\end{document}